\documentclass[12pt]{article}
\usepackage{graphicx}
\usepackage{bm}% bold math
\usepackage{hyperref}
\usepackage{color}

\usepackage{amsmath,amsfonts,amssymb}
\usepackage[left=2.5cm,right=2.5cm,top=2.5cm,bottom=2.5cm,a4paper]{geometry}

\begin{document}

%%%%%%%%%%%%%%%%%%%%%%%%%%%%%%%%%%%%%%%%%%%

\def\o{\over}
\newcommand{\gsim}{ \mathop{}_{\textstyle \sim}^{\textstyle >} }
\newcommand{\lsim}{ \mathop{}_{\textstyle \sim}^{\textstyle <} }
\newcommand{\vev}[1]{ \left\langle {#1} \right\rangle }
\newcommand{\bra}[1]{ \langle {#1} | }
\newcommand{\ket}[1]{ | {#1} \rangle }
\newcommand{\EV}{ \text{eV} }
\newcommand{\KEV}{ \text{keV} }
\newcommand{\MEV}{ \text{MeV} }
\newcommand{\GEV}{ \text{GeV} }
\newcommand{\TEV}{ \text{TeV} }
\newcommand{\1}{\mbox{1}\hspace{-0.25em}\mbox{l}}
\newcommand{\headline}[1]{\noindent{\bf #1}}
\def\diag{\mathop\text{diag}\nolimits}
\def\Spin{\mathop\text{Spin}}
\def\SO{\mathop\text{SO}}
\def\O{\mathop\text{O}}
\def\SU{\mathop\text{SU}}
\def\U{\mathop\text{U}}
\def\Sp{\mathop\text{Sp}}
\def\SL{\mathop\text{SL}}
\def\tr{\mathop\text{tr}}
\def\mpl{M_\text{Pl}}

\def\dd{\mathrm{d}}
\def\ff{\mathrm{f}}
\def\BH{\text{BH}}
\def\inf{\text{inf}}
\def\ev{\text{evap}}
\def\eq{\text{eq}}
\def\SM{\text{sm}}
\def\Mpl{M_\text{Pl}}
\def\GeV{\text{GeV}}
\newcommand{\Red}[1]{\textcolor{red}{#1}}
\newcommand{\vs}{\vspace{50pt}}

%%%%%%%%%%%%%%%%%%%%%%%%%%%%%%%%%%%%%%%%%%%%%%%%%%%%%%%%%%%%%%%
\begin{titlepage}
\baselineskip 8mm
\begin{center}

%\hfill  \\
\hfill \hfill  IPMU-15-0178 \\
%\hfill \today

\vspace{2cm}
{\Large\bf 
Cosmological  Problems of the String Axion 
Alleviated by High Scale SUSY of $m_{3/2}$ $\simeq$ 10\,-100 TeV }

\vspace{2.0cm}
{\bf Masahiro Kawasaki}$^{(a, b)}$
{\bf Tsutomu T. Yanagida}$^{( b)}$
and
{\bf Norimi Yokozaki}$^{( c)}$

\vspace{1.0cm}
{\it
$^{(a)}${ICRR, University of Tokyo, Kashiwa, Chiba 277-8582, Japan}\\
$^{(b)}${Kavli IPMU (WPI), UTIAS, University of Tokyo, Kashiwa, Chiba 277-8583, Japan} \\
$^{(c)}${Istituto Nazionale di Fisica Nucleare, Sezione di Roma,
Piazzale Aldo Moro 2, I-00185 Rome, Italy}
}

\vspace{2.0cm}
\baselineskip 5.7 mm

%{\bf Abstract}

\abstract{%
% To be written. 
The string axion may provide the most attractive solution to the strong CP problem in QCD.
However, the axion energy density easily exceeds the dark matter density in the present universe due to a large decay constant around $10^{16}$ GeV, unless the initial value of the axion field is fine-tuned.
We show that this problem is alleviated if and only if the SUSY particle mass scale
is  10\,-100\,TeV,
since the decay of the saxion can produce a large enough amount of  entropy after the QCD phase transition,
not disturbing the BBN prediction. 
The saxion decay also produces a large number of the lightest SUSY particles (LSPs). 
As a consequence, $R$-parity needs to be violated to avoid the overproduction of the LSPs.
%it is necessary that $R$-parity is violated. 
%the necessary condition is $R$-parity violation.
%
%
%
The saxion field can be stabilized with relatively simple K\"{a}hler potentials, not inducing a too large axion dark radiation. 
Despite the large entropy production, the observed baryon number is explained by the Affleck-Dine mechanism.
Furthermore, the constraint from  
isocurvature perturbations is relaxed,
and the Hubble constant during inflation can be as large as several$\,\times\,10^{10}$\,GeV.
}

\end{center}
\end{titlepage}
\setcounter{footnote}{0}

\baselineskip 5.7mm

%%%%%%%%%%%%%%%%%%%
%---------------SECTION-------------------%

%%%%%%%%%%%%%%%%%%%%%%%%%%%%%%%%%%%%%%%%%%%%%%%%%%%%%%%%%%%%%%%%%
\section{Introduction}
%%%%%%%%%%%%%%%%%%%%%%%%%%%%%%%%%%%%%%%%%%%%%%%%%%%%%%%%%%%%%%%%%
The appearance of an axion in string theories~\cite{Conlon:2006tq, Svrcek:2006yi, Choi:2006za} is extremely attractive, since it may provide the most plausible solution to the strong 
CP problem~\cite{Peccei:1977hh,Weinberg:1977ma,Wilczek:1977pj}.
The axion decay constant $f_a$ is predicted as $f_a \sim 10^{16}$ GeV in many string axion models, which is, however, several orders of magnitude higher than the standard cosmological bounds $f_a \le 10^{12}$ GeV~\cite{Preskill:1982cy,Abbott:1982af,Dine:1982ah}. This upper-bound is obtained for the 
cosmological axion energy density not to exceed the observed dark matter density in the present universe. 
Therefore, one has to finely tune the initial value of the axion field in order to suppress the axion energy density sufficiently.

However, it was pointed out a long time ago in Ref.~\cite{Kawasaki:1995vt} that the above problem can be alleviated if we have a certain amount of entropy production  after the QCD phase transition. As a result, the axion energy density can be consistent with the observed dark matter density without a fine-tuning. We may identify the axion as a dominant component of the cold dark matter in the present universe. 
In addition, 
the upper bound of the Hubble constant during inflation from the isocurvature perturbations is somewhat relaxed as
$H_\text{inf} \lesssim {\rm several} \times 10^{10}$ GeV, which is one order of magnitude larger than the bound without the entropy production~\cite{Kawasaki:1997ct}.

In this letter, we show that the required and consistent entropy production in the late time of the early universe is indeed  provided by the saxion decay if and only if the SUSY particle mass scale is around 100 TeV, without spoiling the  successful prediction of the Big-Bang Nucleosynthesis (BBN).
The axion dark radiation accompanied by the saxion decay, which may be problematic~\cite{dark_rads}, can be suppressed with relatively simple K\"{a}hler potentials, consistent with stabilization of the saxion field. 
%As a consequence, we do not need a fine tuning of the initial axion amplitude and 
%
%

Since a large number of the lightest SUSY particles (LSPs) are produced from the saxion decay, 
we need $R$-parity violation so that the LSPs never exceed the dark matter density in the present universe.
In this case, there is no astrophysical constraint on the LSP such as wino. 
The high scale supersymmetry (SUSY) may be realized by the pure gravity mediation~\cite{puregm}, 
predicting the wino mass in the range of 0.1-1\,TeV.  
Thus, wino searches for the mass $\le 1$ TeV at the LHC  is highly motivated.  
Note that the required size of the $R$-parity violation to avoid cosmological constraints is tiny, and hence, the $R$-parity violation is negligible at the collider time scale in most cases.
%From the above requirement of $H_{\rm inf}$, the present scenario also predicts the tensor to scalar ratio $r\le 10^{-4}$ which is very challenging to be tested in the near future observations.

%%%%%%%%%%%%%%%%%%%%%%%%%%%%%%%%%%%%%%%%%%%%%%%%%%%%%%%%%%%%%%%%%%%%%%%%%%
\section{Axion density and isocurvature perturbations}
\label{sec:isocurvature}
%%%%%%%%%%%%%%%%%%%%%%%%%%%%%%%%%%%%%%%%%%%%%%%%%%%%%%%%%%%%%%%%%%%%%%%%%%
%%%%%%%%%%%%%%%%%%%%%%%%%%%%%%%%%%%%%%%%%%%%%%%%%%%%%%%%%%%%%%%%%%%%%%%%%%
\subsection{Without entropy production}
%%%%%%%%%%%%%%%%%%%%%%%%%%%%%%%%%%%%%%%%%%%%%%%%%%%%%%%%%%%%%%%%%%%%%%%%%%

In this section we briefly review the cosmic density of the string axion and how stringent constraint on the Hubble parameter during inflation is derived from the isocurvature perturbations of the axion density.
First let us consider the case without entropy production.
In this case, the string axion density in the present universe is given by~\cite{axion_cdm} 
\begin{equation}
   \Omega_a h^2 = 1.04\times 10^{4} \Theta^2
   \left(\frac{f_a}{10^{16}\text{GeV}}\right)^{1.19},
   \label{eq:density_wo_entropy}
\end{equation}
where $\Theta$ is misalignment angle and $h$ is the present Hubble parameter in units of {100\,km/s/Mpc}, 
and the initial amplitude of the axion field is $f_a \Theta $. 
The axion density should not exceed the present dark matter density $\Omega_{\rm DM} h^2 \simeq 0.12$, from which we obtain
\begin{equation}
   \Theta < 3.4\times 10^{-3}
   \left(\frac{f_a}{10^{16}\text{GeV}}\right)^{-0.59} .
\end{equation}
Thus, the misalignment angle $\Theta$ should be very small for the string axion with $f_a\simeq 10^{16}$~GeV.

During inflation the axion field $a(=\Theta f_a)$ acquires fluctuations as
\begin{equation}
   \delta a =f_a \delta\Theta \simeq \frac{H_{\text{inf}}}{2\pi}  ,
\end{equation}
where $H_{\text{inf}}$ is the Hubble parameter when the pivot scale ($k_*=0.05\text{ Mpc}^{-1}$) left the horizon during inflation.
This leads to the isocurvature perturbations~\cite{Linde:1984ti,Seckel:1985tj,Lyth:1989pb,Turner:1990uz,Linde:1991km} whose power spectrum is given by
\begin{equation}
   P_{\text{iso}} \simeq 4 \left(\frac{\Omega_a h^2}{\Omega_{\rm DM} h^2}\right)^2
   \left(\frac{\delta a}{a}\right)^2 
   \simeq  \left(\frac{\Omega_a h^2}{\Omega_{\rm DM} h^2}\right)^2
   \left(\frac{H_{\text{inf}}}{\pi f_a \Theta}\right)^2 .
   \label{eq:power_spec_iso}
\end{equation}
Since the observed CMB anisotropies are consistent with the pure adiabatic density perturbations, the isocurvature perturbations are stringently constrained~\cite{Ade:2015lrj} as
\begin{equation}
   \beta_{\text{iso}} \equiv \frac{P_{\text{iso}}}{P_{\text{ad}}}
   < 0.037 .
   \label{eq:iso_const}
\end{equation}
If the axion is dark matter, the constraints on the isocurvature perturbations leads to the upper bound on the Hubble during inflation,
\begin{equation}
   H_{\text{inf}} \lesssim 9.6 \times 10^{8}\text{ GeV}
   \left(\frac{f_a}{10^{16}\text{ GeV}}\right)^{0.41}.
\end{equation}
Here we have used $P_\text{ad}(k_*)\simeq 2.2\times 10^{-9}$~\cite{Ade:2015lrj} and 
Eq.~(\ref{eq:density_wo_entropy}) with $\Omega_a \simeq \Omega_{\rm DM}$.
%with $\Theta \simeq 3.4 \times 10^{-3} (f_a/10^{16}\text{GeV})^{-0.59}$.

%%%%%%%%%%%%%%%%%%%%%%%%%%%%%%%%%%%%%%%%%%%%%%%%%%%%%%%%%%%%%%%%%%%%%%%%%%
\subsection{With entropy production}
%%%%%%%%%%%%%%%%%%%%%%%%%%%%%%%%%%%%%%%%%%%%%%%%%%%%%%%%%%%%%%%%%%%%%%%%%%

If entropy production takes place during or after the start of the axion oscillation the cosmic axion density is diluted. 
For entropy production with reheating temperature $T_R$ the axion density
is written as~\cite{Kawasaki:1995vt}
\begin{equation}
   \Omega_a h^2 = 5.3 \left(\frac{T_R}{\text{MeV}}\right)
   \left(\frac{f_a \Theta}{10^{16}\text{ GeV}}\right)^2 .
   \label{eq:density_w_entropy}
\end{equation}
%{\bf 
Here it is assumed that the entropy is produced by decays of some heavy particles (saxion in our case) which dominate the universe
for some time before the axion starts to oscillate.
In order not to spoil the BBN the reheating temperature
should be larger than $O(1)$~MeV~\cite{Kawasaki:1999na}. 
 
The isocurvature perturbations are given by Eq.~(\ref{eq:power_spec_iso}).
From the constraint (\ref{eq:iso_const}) we obtain
 \begin{equation}
    H_{\text{inf}} \lesssim 0.64\times 10^{10}~\text{GeV}
    \left(\frac{f_a\Theta}{10^{16}\text{ GeV}}\right)^{-1}
    \left(\frac{T_R}{\text{MeV}}\right)^{-1} .
 \end{equation}
For the dark matter axion ($\Omega_a h^2 \simeq 0.12$), $f_a \Theta$ is given by
\begin{equation}
   f_a \Theta = 1.51\times 10^{15}~\text{GeV}
   \left(\frac{T_R}{\text{MeV}}\right)^{-1/2} ,
   \label{eq:angle_w_entropy}
\end{equation}
which leads to the upper bound on the Hubble parameter during inflation, 
\begin{equation}
   H_{\text{inf}} \lesssim 4.3\times 10^{10}~\text{GeV}
   \left(\frac{T_R}{\text{MeV}}\right)^{-1/2} .
\end{equation}
With entropy production, therefore, the constraint on the Hubble parameter during inflation becomes more than one order of magnitude milder.
Furthermore, since the required misalignment angle $\Theta$ is $\sim O(0.1)$, we do not need unnatural fine tuning.

%%%%%%%%%%%%%%%%%%%%%%%%%%%%%%%%%%%%%%%%%%%%%%%%%%%%%%%%%%%%%%%%%%%%%%%%%%
\section{Saxion decays}
\label{sec:saxion}
%%%%%%%%%%%%%%%%%%%%%%%%%%%%%%%%%%%%%%%%%%%%%%%%%%%%%%%%%%%%%%%%%%%%%%%%%%

\subsection{BBN and Entropy production}
In this section we show that the entropy production required for diluting the string axion is provided by the saxion decay.
During inflation the saxion generally has a large field value $\sim M_p$ ($M_p$\,: reduced Planck mass $\simeq 2.4\times 10^{18}~\text{GeV}$).
After inflation, when the Hubble parameter becomes equal to the saxion mass, the saxion starts oscillation.
Since the initial oscillation amplitude is order of $M_p$, the saxion likely dominates the universe.
So when the saxion decays it produces large entropy.

The saxion decays into gauge bosons (and gauginos) through the following operator:
\begin{eqnarray}
\mathcal{L} = \int d^2 \theta \left(\frac{1}{4g_i^2} + \frac{\mathcal{A} \sqrt{2}}{32\pi^2 f_a} \right) (\mathcal{W}^{c \, \alpha})_i (\mathcal{W}^c_\alpha)_i + h.c.,
\end{eqnarray}
where $\mathcal{A}$ is an axion chiral superfield, and is canonically normalized. Here, $g_1$, $g_2$ and $g_3$ denote the  gauge coupling constants of  $U(1)_Y$, $SU(2)_L$ and $SU(3)_c$, and 
$\mathcal{W}_i$ is a field strength superfield. 
The axion $a$ and saxion $\sigma$ are contained in a scalar component of $\mathcal{A}$ as
\begin{eqnarray}
\mathcal{A} |_{\theta=\bar \theta=0} = \frac{1}{\sqrt{2}} ( \sigma + i a).
\end{eqnarray}
The Lagrangian contains 
\begin{eqnarray}
\mathcal{L} \ni -\frac{g_3^2 }{32\pi^2 f_a} \left( \sigma F_{\mu \nu}^c F^{\mu \nu\,c}  + \frac{\epsilon^{\mu \nu \kappa \lambda}}{2}  a F_{\kappa \lambda}^c F_{\mu \nu}^c \right),
\end{eqnarray}
where $F_{\mu \nu}^c$ is the gluon field strength tensor. Here, the gauge fields are canonically normalized.
Then, the decay rate of the saxion into gluons is given by
\begin{equation}
   \Gamma(\sigma \to gg) = \frac{\alpha_s^2}{32\pi^3}\frac{m_{\sigma}^3}{f_a^2},
\end{equation}
where $\alpha_s=g_3^2/(4\pi)$ and $m_{\sigma}$ is the saxion mass.

If the gluino mass $m_{\tilde g}$ is smaller than $m_{\sigma}/2$, 
the saxion also decays into the gluinos with a similar ratio, $\Gamma(\sigma \to \tilde g \tilde g) \simeq (m_{3/2}/ m_{\sigma})^2\Gamma(\sigma \to  g  g)$, neglecting $(m_{\tilde g}/m_{\sigma})$ corrections.\footnote{
It is assumed that $\left<F_\mathcal{A}\right>=0$, and the superpotential does not depend on $\mathcal{A}$, which may be necessary to keep the shift-symmetry unbroken. 
}
Here, $m_{3/2}$ is a gravitino mass.
Then, the reheating temperature $T_{R}$ is written as 
\begin{align}
      T_{R} & \simeq \left(\frac{\pi^2 g_*}{90}\right)^{-1/4}
   \left(\Gamma_{\rm total} M_p\right)^{1/2} ,\\
          &  \simeq 11~\text{MeV} \eta
              \left(\frac{m_{\sigma}}{100~\text{TeV}}\right)^{3/2}
              \left(\frac{f_a}{10^{16}~\text{GeV}}\right)^{-1},
\end{align}
where $\Gamma_{\rm total} \simeq \eta^2 \, \Gamma (\sigma \to gg)$ is assumed.
Here $g_* (\simeq 43/4)$ is the relativistic degrees of freedom,
$\alpha_s(\mu_R=50\,{\rm TeV})=0.07$ ($\mu_R$ is a renormalization scale),\footnote{
Here, the one-loop renormalization group equation is used:
\begin{eqnarray}
\alpha_s(\mu_R)^{-1} = \alpha_s(m_Z)^{-1} + \frac{7}{2\pi}\ln \frac{m_{\tilde g}}{m_Z} +   \frac{5}{2\pi}\ln \frac{\mu_R}{m_{\tilde g}}, \ 
{\rm with} \ \, m_{\tilde g}=1\, {\rm TeV}.
\end{eqnarray}
}
 and $\eta=1, (\,\eta=\sqrt{1 + m_{3/2}^2/m_{\sigma}^2}\,)$
 in the case that the saxion does not decay (decays) into the gluinos.

Requiring that the decay of the saxion do not disturb the BBN prediction, i.e. $T_R > \mathcal{O}
({\rm MeV})$,  the lower-bound on the saxion mass is obtained as
\begin{eqnarray}
T_R > 1\,(5)\, {\rm MeV} \to \ \ m_{\sigma} \gtrsim 20\, (60)\, {\rm TeV} \left(\frac{f_a}{10^{16}\,{\rm GeV}}\right)^{2/3} \eta^{-2/3}.
\end{eqnarray}
On the other hand, the produced entropy from the saxion decay is reduced for larger $m_{\sigma}$, 
and hence the SUSY particle mass scale is bounded from above as (see Eq.\,(\ref{eq:density_w_entropy})) 
\begin{eqnarray}
\Omega_a h^2 \lesssim 0.12 \to m_{\sigma} \lesssim 35\, {\rm TeV} \left(\frac{f_a}{10^{16}\,{\rm GeV}}\right)^{-2/3} \left(\frac{0.1}{\Theta}\right)^{4/3} \eta^{-2/3}.
\end{eqnarray}
Therefore, the gravitino mass is expected to be in a range of 10-100\,TeV, depending on a relation between $m_{3/2}$ and $m_{\sigma}$.

Before closing this subsection, let us estimate the dilution factor due to the entropy production
by the saxion decay.
Assuming that the saxion with initial amplitude $\sigma_0$ starts to oscillate before the inflaton decays,
%(After the inflaton decay,) 
the ratio of the saxion density $\rho_s$ to the entropy density $s_i$
due to the inflaton decay is
\begin{equation}
    \frac{\rho_s}{s_i} \simeq \frac{1}{8} T_{IR}\left(\frac{\sigma_0}{M_p}\right)^2 ,
\end{equation}
where $T_{IR}$ is the reheating temperature of the inflaton decay.
% and $\sigma_0$ is the initial field valueof the saxion.
%Here we assume that the axion starts to oscillate  before the inflaton decays.
When the saxion decays, its energy density is transferred to the entropy density $s_f$, 
and the dilution factor $\Delta =s_f/s_i$ amounts to
\begin{equation}
   \Delta \simeq     \frac{1}{6}\,  \frac{T_{IR}}{T_{R}}\left(\frac{\sigma_0}{M_p}\right)^2. \label{eq:dilution_fac}
\end{equation}
Therefore, the saxion produces a huge dilution factor $\sim 10^{10}$ for $T_{IR} \sim 10^{8}$~GeV
and $\sigma_0 \sim M_p$.

%%%%%%%%%%%%%%%%%%%%%%%%%%%%%%%%%%%%%%%%%%%%%%%%%%%%%%%%%%%%%%%%%%%%%%%%%%%%
\subsection{The dark radiation and saxion stabilization} \label{sec:stab}
%%%%%%%%%%%%%%%%%%%%%%%%%%%%%%%%%%%%%%%%%%%%%%%%%%%%%%%%%%%%%%%%%%%%%%%%%%%%
In this scenario, the saxion also decays into axions, producing a dark radiation. If this decay rate is too large, 
the produced dark radiation may exceed the cosmological bound~\cite{dark_rads}; therefore, the saxion decay into axions should be suppressed.
 
The CMB observation by Planck~\cite{Ade:2015xua} gives the constraint on the dark radiation as
\begin{equation}
   \Delta N_{\nu} < 0.4,
\end{equation}
where the energy density of the dark radiation is conventionally expressed by the extra effective neutrino number $\Delta N_{\nu}$.
In our case,
\begin{equation}
   \Delta N_{\nu} = \frac{43}{7}\left(\frac{g_*}{43/4}\right)^{4/3}
   \frac{ {\rm Br} (\sigma\rightarrow 2a)}{1-{\rm Br}(\sigma\rightarrow 2a)}.
\end{equation}
Thus, the branching ratio into axions should be
\begin{equation}
   {\rm Br}(\sigma\rightarrow 2a) < 0.06.
\end{equation}

Since the branching ratio depends on how to stabilize the saxion field, 
we consider examples of $N=1$ supergravity models  
as a `bottom-up' approach to the string theoretic axion. 
In the rest of this section, we take the unit of $M_p=1$, unless it is explicitly written.

\paragraph{Example 1}
First, the following K\"{a}hler potential is considered as a simple example,
\begin{eqnarray}
K =   \frac{c_2}{2} x^2 + \frac{c_3}{6} x^3  ,
\end{eqnarray}
where $x=\mathcal{A} + \mathcal{A}^\dag$, $c_2$ and $c_3$ are real constants.
It is assumed that the superpotential, $W$, is not a function of $\mathcal{A}$, since otherwise the shift-symmetry is broken and the strong CP-problem is not solved anymore; therefore, we assume $W=\mathcal{C}$ with 
$m_{3/2}^2 = e^{\left<K\right>} |\mathcal{C}|^2$.
The scalar potential is given by
\begin{eqnarray}
V = e^{K}  \left[ 
\frac{(c_2 x + c_3 x^2/2)^2}{c_2 + c_3 x} |W|^2  -3 |W|^2 + \dots \right]  ,
\end{eqnarray}
where $\dots$ denotes contributions from the SUSY breaking.\footnote{
Because of the SUSY breaking effect, the no-go theorem~\cite{Conlon:2006tq} is avoided.
}
With this potential, $x$ is stabilized at $\left<x\right>=0$ for $c_2>0$ and the vanishing cosmological constant.
Then, the canonically normalized field $\mathcal{A}$ is obtained by rescaling $ c_2^{1/2} \mathcal{A} \to \mathcal{A}$, and 
the saxion mass  is found to be $m_{\sigma}=2 m_{3/2}$.\footnote{
As long as the minimum of the potential is determined by $(\partial K/\partial x)=0$, 
the predicted saxion mass is always $m_{\sigma}=2 m_{3/2}$.
} 
Now, the K\"{a}hler potential becomes
\begin{eqnarray}
K =   \frac{1}{2} x^2 +   \left(c_3 c_2^{-3/2} \right) \frac{x^3}{6}.
\end{eqnarray}
The second term induces the saxion decay into two axions:
\begin{eqnarray}
\Gamma (\sigma \to 2 a)  \simeq (c_3^2 c_2^{-3}) \frac{1}{32\pi}¡¡\frac{m_{\sigma}^3}{M_p^2}.
\end{eqnarray}
Then, the branching ratio is
\begin{eqnarray}
{\rm Br}(\sigma \to 2 a) &\simeq& 
\frac{\Gamma(\sigma \to 2 a)}{\Gamma(\sigma \to gg) + \Gamma(\sigma \to \tilde g \tilde g)} \nonumber \\
& \simeq&  0.03 \cdot (c_3^2 c_2^{-3}) \left(\frac{f_a}{10^{16}\,{\rm GeV}} \right)^2,
\end{eqnarray}
which is small enough for $|c_3/c_2^{3/2}| \lesssim 1.4 $.
The axino mass $m_{\tilde a}$ equals to $m_{3/2}$, which can be seen by the K\"{a}hler transformation, 
$K \to K - A^2/2 -A^{\dag\,2}/2, \,\, W\to e^{A^2/2}W$.
Note that the axino (gravitino) decays into gluon and gluino, and is not stable 
if the mass of the axino (gravitino) is larger than the gluino mass.

\paragraph{Example 2}
Next, we consider a case that the saxion is stabilized by SUSY breaking effects. 
%This setup is similar to the KKLT scenario~\cite{KKLT}, but without a superpotential for $\mathcal{A}$.
The K\"{a}hler potential is 
\begin{eqnarray}
K =  - n \ln \left( 1 + c_1 x \right) - k_Z x ^2 |Z|^2 + |Z|^2, \ \ W=w(Z) +{\mathcal{C}},
\end{eqnarray}
where $Z$ is a SUSY breaking field, and $\left<Z\right> \simeq 0$ and $w(0) \simeq 0$ are assumed, which may be protected by a symmetry of a hidden sector.
This symmetry forbids tree-level gaugino masses from $F_Z$, 
but it is expected that there exists a contribution from anomaly mediation~\cite{amsb}.
The scalar potential is given by
\begin{eqnarray}
V = e^{K} \Bigl( |m_{3/2}|^2 (n-3) 
+ \left|\frac{\partial w}{\partial Z} \right|^2 \frac{1}{1- k_Z x^2} \Bigr),
\end{eqnarray}
and $x$ is stabilized at the origin, $\left<x\right>=0$ for $V|_{\rm min}=0$. The saxion mass is given by
\begin{eqnarray}
m_{\sigma}^2 = \frac{4 k_Z (3-n)} {n c_1^2} m_{3/2}^2, 
\ {\rm with} \ \, \frac{\partial^2 K}{\partial x^2} \Bigr|_{x=\left<x\right>} =  n c_1^2,
\end{eqnarray}
and $m_{\tilde a}=m_{3/2}$. 
The $F$-term of $\mathcal{A}$ is 
\begin{eqnarray}
\left< F_{\mathcal{A}}  \right> \simeq -\frac{m_{3/2}}{c_1}.
\end{eqnarray}
Because of $|\left<F_{\mathcal{A}}\right> (\partial^2 K/\partial x^2)_{x=\left<x\right>}^{1/2}| \sim m_{3/2} $, the gauginos are likely to be heavy  as  the gravitino.\footnote{
The gaugino mass arises from a operator,
\begin{eqnarray}
\mathcal{L} \ni \int d^2 \theta   \, g_i^2 \,  \frac{ \sqrt{2} (\partial^2 K/\partial x^2)^{1/2} \left<F_{\mathcal{A}} \right>  \theta^2 }{32 \pi^2 f_a} (\mathcal{W}^{c \, \alpha})_i (\mathcal{W}^c)_i + h.c.,
\end{eqnarray}
where the field strength superfields are canonically normalized. 
}
After canonically normalizing the field $\mathcal{A}$, 
the saxion-axion-axion coupling is extracted from
\begin{eqnarray}
\int d^4 \theta \frac{1}{6} \left[\frac{\partial^3 K}{\partial x^3} \left(\frac{\partial^2 K}{\partial x^2}\right)^{-3/2} \right]_{x=0} x^3 \ni  -{\rm sign}(c_1)
 \sqrt{\frac{2}{n}} \sigma \partial_{\mu} a \partial^{\mu} a .
\end{eqnarray}
Therefore, the saxion decay rate into the axions and its branching ratio are
\begin{eqnarray}
\Gamma(\sigma \to 2a) &=& \frac{1}{8\pi} \frac{1}{n} \frac{m_{\sigma}^3}{M_p^2}, \ \ \nonumber \\
{\rm Br}(\sigma \to 2a) &\simeq& \frac{\Gamma(\sigma \to 2a)}{\Gamma(\sigma \to 2g)} \simeq 0.14 \cdot \frac{1}{n} \left( \frac{f_a}{10^{16} {\rm GeV}}\right)^2,
\end{eqnarray}
which is marginal for e.g. $f_a=9\cdot 10^{15}$ GeV and $n=2$. 
%One may need a smaller $f_a$, e.g. $f_a = 9 \cdot 10^{15}$\,GeV for $n=2$.

\paragraph{Example 3}
Here, we consider an example with a logarithmic K\"{a}hler potential,
where $x$ does not couple to the SUSY breaking field. The K\"{a}hler potential and superpotential are 
\begin{eqnarray}
K =  - n \ln \left[x - d_2 x^2/2 \right], \ \ W=\mathcal{C}.
\end{eqnarray}
For $n=3$ and $d_2=0$, this K\"{a}hler potential takes the no-scale form, 
and for $n=1$ and $d_2=0$, it takes the same form as the K\"{a}hler potential for the dilaton. 
However, we consider more general cases here.
%{\bf The non-zero $d_2$ may arise from higher order corrections.} 
As in the former two examples, it is assumed that the SUSY is broken and the cosmological constant vanishes.
The scalar potential is written as
\begin{eqnarray}
V = e^{K} \left[\frac{2n(1- d_2 x)^2}{2 - 2 d_2 x + d_2^2 x^2} |W|^2 -3 |W|^2 
+ \dots
\right] .
\end{eqnarray}
The minimum of the potential is found with $\left<x\right>=1/d_2$. 
For $d_2 > 0$ the saxion mass is $m_{\sigma}=2m_{3/2}$, $m_{\tilde a}=m_{3/2}$, and $\left<F_{\mathcal{A}}\right>=0$, 
where $(\partial^2 K/ \partial x^2)_{x=\left<x\right>}=(2 n)  d_2^2 $ is used.
%

%\begin{eqnarray}
%m_{\sigma}^2=  2^{2+n} d_2^{\, n}\, m_{3/2}^2,
%end{eqnarray}
%where $m_{\sigma} \approx m_{3/2}$ for $d_2 \simeq 0.13$ and $n=1$. 
%
The decay rate of the saxion into the two axions is highly suppressed, since $(\partial^3 K/ \partial x^3) =0$ at the minimum:
this model can easily avoid the constraint from the dark radiation.
As shown in Appendix \ref{ap:sax},  similar models with $K=\left[-n \ln(x) + d_2'x^2/2\right]$, $\left[-n  \ln(x- d_3 x^3/6)\right]$ and $\left[-n \ln(x) + d_3' x^3/6 \right]$ can also avoid the constraint from the dark radiation easily.

%%%%%%%%%%%%%%%%%%%%%%%%%%%%%%%%%%%%%%%%%%%%%%%%%%%%%%%%%%%%%%%%%%%%%%%%%%
\section{Dark matter}
\label{sec:dark_matter}
%%%%%%%%%%%%%%%%%%%%%%%%%%%%%%%%%%%%%%%%%%%%%%%%%%%%%%%%%%%%%%%%%%%%%%%%%%
In our scenario the string axion can be dark matter of the universe.
If the $R$-parity is conserved, the lightest supersymmetric particle (LSP) is also a candidate for dark matter.
Since thermal relic particles are diluted away by the entropy production due to the saxion decay (see Eq.\,(\ref{eq:dilution_fac})), 
we only need to consider the LSPs produced in the saxion decay. 
The LSP density is given by 
\begin{align}
   \frac{\rho_{\text{LSP}}}{s} &\simeq 
   \frac{3 m_{\text{LSP}}(\rho_\sigma/m_{\sigma})}{4\rho_\sigma/T_R} 
   B_{\text{LSP}}
   \simeq 
   \frac{3 m_{\text{LSP}}}{4 m_\sigma}T_R B_{\text{LSP}} \nonumber \\
   & \simeq 0.75\times 10^{-6}~\text{GeV} 
   \left(\frac{m_\sigma}{100~\text{TeV}}\right)^{-1}
   \left(\frac{m_\text{LSP}}{100~\text{GeV}}\right)
   \left(\frac{T_R}{\text{MeV}}\right) B_{\text{LSP}} ,
\end{align}
where $s$ is the entropy density, $\rho_\sigma$ is the saxion density, $m_{\text{LSP}}$ is the LSP mass and $B_{\text{LSP}}$ is the branching ratio of the saxions into the LSPs including subsequent decays of SUSY particles into the LSPs. 
The above equation shows that the LSP density is too large for $B_{\text{LSP}}\sim 1$. 
(Notice that $\rho_c/s \simeq 3.6\times 10^{-9}h^2~\text{GeV}$ where $\rho_c$ is the critical density.)   
In fact, if the LSP is the wino as predicted in the pure gravity mediation, $B_{\tilde{W}} \sim  1$.
%~\cite{Endo:2006ix}, 
Thus, the stable LSP is not consistent 
with the present DM density and hence we need a $R$-parity violation to make the LSP unstable.
%with the present model and LSPs should decay due to the $R$-parity violation.

%%%%%%%%%%%%%%%%%%%%%%%%%%%%%%%%%%%%%%%%%%%%%%%%%%%%%%%%%%%%%%%%%%%%%%%%%%
\section{Baryogenesis}
\label{sec:baryogenesis}
%%%%%%%%%%%%%%%%%%%%%%%%%%%%%%%%%%%%%%%%%%%%%%%%%%%%%%%%%%%%%%%%%%%%%%%%%%

Since the entropy production by the saxion dilutes pre-existing baryon number, a large baryon asymmetry should exit before the entropy production in order to account for the present baryon number density of the universe.
Here, we consider the Affleck-Dine (AD) baryogenesis~\cite{AD} which can produce baryon number efficiently. 
For simplicity, we assume that the AD field has the initial field value $\phi_0$ and there are no non-renormalizable terms so that the AD field stays at $\phi_0$ until it starts to oscillate. 
When the AD field starts oscillation, the ratio of the created baryon number to the total cosmic density $\rho$ is
\begin{equation}
   \left. \frac{n_b}{\rho} \right|_{\text{osc}}
   \simeq \frac{1}{6}\,\epsilon\frac{m_\phi \phi_0^2}{m_\phi^2 M_p^2}  
   = \frac{1}{6}\, \frac{\epsilon}{m_\phi}\left(\frac{\phi_0}{M_p}\right)^2 ,
\end{equation}
where $m_\phi$ is the mass of the AD field and $\epsilon\, (< 1)$ is the efficiency parameter which denotes the largeness of the kick in the phase direction at the start of the oscillation. 
When the universe is dominated by the saxion, the total density is nearly equal to the saxion density, $\rho \simeq \rho_\sigma$.
In addition, $n_b/\rho_{\sigma}$ remains constant until the saxion decay.
Thus the baryon-to-entropy ration is given by
\begin{equation}
   \frac{n_b}{s} 
   \simeq \frac{3}{4}\,\frac{n_b}{\rho_\sigma/T_R}
   \simeq \frac{1}{8}\epsilon \frac{T_R}{m_\phi}
   \left(\frac{\phi_0}{M_p}\right)^2 
   \sim 10^{-9}\epsilon 
   \left(\frac{T_R}{\text{MeV}}\right)
   \left(\frac{m_\phi}{100~\text{TeV}}\right)^{-1}
   \left(\frac{\phi_0}{M_p}\right)^2 .
\end{equation}
Thus, sufficient baryon asymmetry is produced if $\phi_0\sim M_p$ and 
$\epsilon \sim 10^{-1}$.

\section{Conclusion and discussion}

In this letter we have shown that the string axion is consistent with the standard cosmology in the presence of the large entropy production due to the decay of the saxion of mass 10-100\,TeV.
In this scenario, the string axion is the dark matter of the universe for misalignment angle $\Theta \sim O(0.1)$. 
The axion isocurvature perturbations are small enough if the Hubble parameter during inflation is less than  $\text{several}\times 10^{10}$~GeV, which is more than one order of magnitude milder than the constraint without entropy production. 
Furthermore, the baryon number of the universe is provided by the Affleck-Dine mechanism.
Although too many LSPs are produced by the saxion decay, 
they do not contribute to the energy density of the present universe. 
This is because they are unstable due to the $R$-parity violation.
%the overproduction can be avoided by the $R$-parity violation.
%
%Overproduction of the LSPs in the saxion decay is avoided by the $R$-parity violation. 
%The saxion decay produces not only large entropy but also significant dark radiation which may be detected in future observations.

The large saxion mass requires the  high SUSY breaking scale as $m_{3/2}=10$\,-$100$\,TeV.
This may be realized by the pure gravity mediation which predicts that the LSP is  wino of mass 0.1-1 TeV.
In the present scenario the wino is not dark matter due to $R$-parity violation and hence it evades astrophysical~\cite{Bhattacherjee:2014dya,Ackermann:2015zua} and cosmological~\cite{Ade:2015xua,Kawasaki:2015yya} constraints, which makes the wino searches at the LHC highly motivated. Since the required size of $R$-parity violation is tiny and negligible at the collider time scale in most cases, the standard strategies for the SUSY searches at the LHC are applicable.
The gluino mass is likely to be smaller than 3 TeV,\footnote{
The prediction of the gaugino masses in anomaly mediation changes if $F_{\mathcal{A}}$ has a VEV of $\mathcal{O}((
0.01$-$0.1)m_{3/2})$ as shown in Appendix\,\ref{ap:susybreaking}, which can lift up the masses.
} 
since the sufficient entropy production leads to the upper-bound on the saxion mass (i.e. $m_{3/2}$) as shown in Sec.~\ref{sec:saxion}.
Therefore, the gluino can be discovered at the 13-14 TeV LHC.

\section*{Acknowledgements}
This work is supported by Grants-in-Aid for Scientific Research from the Ministry of Education, Culture, Sports, Science, and Technology (MEXT), Japan,
No. 25400248 (M.~K.)
and  No. 26104009 (T.~T.~Y.); 
MEXT Grant-in-Aid for Scientific Research on Innovative Areas No.15H05889 (M.~K.);
Grant-in-Aid No. 26287039 (T.~T.~Y.) from the Japan Society for the Promotion of Science (JSPS); and by the World Premier International Research Center Initiative (WPI), MEXT, Japan (M.~K. and T.~T.~Y.).
The research leading to these results has received funding
from the European Research Council under the European Unions Seventh
Framework Programme (FP/2007-2013) / ERC Grant Agreement n. 279972
``NPFlavour'' (N.\,Y.).

\appendix

\section{Other examples of the saxion stabilization} \label{ap:sax}
Here, we consider another model, where the $F$-term of $\mathcal{A}$  is non-vanishing. The K\"{a}hler potential is, 
\begin{eqnarray}
K =  - n \ln (x) + d_2' x^2/2,
\end{eqnarray}
where $x=\mathcal{A} + \mathcal{A}^\dag$, and $d_2'$ is a  positive constant. Here, we take the unit of $M_p=1$.
%
%The scalar potential is 
%\begin{eqnarray}
%V = e^{K} |m_{3/2} M_p|^2 \frac{(d_1' + x^2 d_2')^2}{-d_1' + x^2 d_2'}
%\end{eqnarray}
For $n>0$, the minimum of $x$ is $\left<x\right>= \sqrt{n/d_2'}$, and 
\begin{eqnarray}
\frac{\partial^2 K}{\partial x^2} = 2 d_2', 
\ \ \frac{1}{\sqrt{2}} \frac{\partial^3 K}{\partial x^3} \left( \frac{\partial^2 K}{\partial x^2} \right)^{-3/2} = -\frac{1}{2 \sqrt{n} }.
\end{eqnarray}
The decay rate of the saxion into the two axions is,
\begin{eqnarray}
\Gamma (\sigma \to 2 a) \simeq \frac{1}{64\pi}\frac{1}{n}\frac{m_\sigma^3}{M_p^2},
\end{eqnarray}
and hence, for $n \gtrsim 1$, the dark radiation may be enough suppressed. The saxion mass is predicted to be
$m_{\sigma}=2m_{3/2}$, $m_{\tilde a}=m_{3/2}$, and $\left<F_{\mathcal{A}}\right>=0$.

\vspace{10pt}
Even if there is a cubic term of $x$ inside a log function,  $\Gamma(\sigma \to 2 a)$ can be suppressed. 
We take
\begin{eqnarray}
K =  - n \ln (x - d_3 x^3/6),
\end{eqnarray}
where the minimum is found as $\left<x\right>=\sqrt{2/d_3}$, and
\begin{eqnarray}
\frac{\partial^2 K}{\partial x^2} = \frac{3}{2}n d_3, \  \, \frac{1}{\sqrt{2}} \frac{\partial^3 K}{\partial x^3} 
\left(\frac{\partial^2 K}{\partial x^2} \right)^{-3/2} = \frac{1}{\sqrt{6 n}}.
\end{eqnarray}
Therefore, $\Gamma(\sigma \to 2a) = m_{\sigma}^3/(96\pi n M_p^2)$, which is small enough.
The saxion mass is $m_{\sigma}=2 m_{3/2}$, $m_{\tilde a}=m_{3/2}$, 
and $F_{\mathcal{A}}$ vanishes at the minimum.
Also, the model with $K=- n \ln (x) + d_3' x^3/6$ ($d_3'>0$) is safe: the dark radiation constraint can be avoided.

\section{An example of a small $F_{\mathcal{A}}$} \label{ap:susybreaking}
We consider the following K\"{a}hler potential and superpotential:
\begin{eqnarray}
K = c_2 x^2/2 + k_Z' x |Z|^2 + |Z|^2, \ \ W=w(Z)+\mathcal{C},
\end{eqnarray}
where the unit of $M_p=1$ is taken, and $\left<Z\right> \simeq 0$ and $w(0) \simeq 0$ are assumed. 
Also, $k_Z' \ll c_2$ is assumed, which may be natural if $x |Z|^2$ arises from a higher order correction.
In this case, the minimum of $x$ is slightly deviated from zero as $\left<x\right>=3k_Z'/(2c_2)$, and the non-vanishing $F$-term arises.
\begin{eqnarray}
\left<F_{\mathcal{A}}\right> = -e^{K/2} \frac{\partial K}{\partial x} \left(\frac{\partial^2 K}{\partial x^2}\right)^{-1} \mathcal{C}^* \simeq -\frac{3 k_Z'}{2 c_2} m_{3/2}. 
\end{eqnarray}
Therefore, the contribution to the gaugino masses is 
\begin{eqnarray}
\left<F_{\mathcal{A}}\right> (\partial^2 K/ \partial x^2)^{1/2} \sim  k_Z'/c_2^{1/2} m_{3/2}, 
\end{eqnarray}
which can be comparably small to those from anomaly mediation.
The masses of the saxion and axino are almost identical to $2m_{3/2}$ and $m_{3/2}$.

\end{document}